# Linear up-conversion of orbital angular momentum

Dong-Sheng Ding, Zhi-Yuan Zhou, Bao-Sen Shi[*], Xu-Bo Zou[*] and Guang-Can Guo

*Key Laboratory of Quantum Information, University of Science and Technology of China, Hefei 230026, China*
[*]*Corresponding author: drshi@ustc.edu.cn*
*xbz@ustc.edu.cn*

We experimentally demonstrated that infrared light imprinted with orbital angular momentum (OAM) was linearly converted into visible light using four-wave mixing (FWM) via a ladder-type configuration in $^{85}$Rb atoms. Simultaneously, we theoretically simulated this linear conversion process, and the theoretical analysis was in reasonable agreement with the experimental results. A large single-photon detuning process was used to reduce the absorption of the atoms to the up-converted light and to avoid pattern formation in the FWM process. The multi-mode image linear conversion process is important for applications including image communications, astrophysics and quantum information. © 2012 Optical Society of America

Optical beams with OAM states include screw topological wavefront dislocations or vortices. Light with OAM has many exciting applications, including optical communications [1-2] and astrophysics [3-5]. Because of the lack of a suitable detector technology in the infrared region, efficient conversion of infrared light with OAM to the visible region with high resolution is very important for these applications. Although the transfer of an OAM state through an optical frequency conversion process in nonlinear crystal and atomic ensemble has been reported [6-8], no linear up-conversion process has been reported for an OAM state from the infrared to the visible region in atoms. The linearity of up-conversion from infrared to visible light may for example be used to improve the angular resolution of astronomical instruments [9]. The linear conversion of OAM may also find possible applications in combination with an entangled image [10], and with the quantum correlated angle [11].

The up-conversion of light from the infrared spectrum to the visible region can be realized with a diamond or a ladder atomic configuration [12-15]. A theoretical scheme for efficient infrared image up-conversion via quantum coherence based on FWM is present in Ref. [12]. In. Ref. [13], more than 50% up-conversion efficiency from infrared light to visible light was obtained in a cold atomic ensemble. However, these experiments were achieved using techniques without spatial resolution. Here, we experimentally demonstrate that infrared light imprinted OAM can be up-converted linearly into the visible region through the FWM process in a ladder-type configuration in $^{85}$Rb atoms. We also theoretically simulate this linear conversion process, and the theoretical analysis is in reasonable agreement with the experimental results. A large single-photon detuning process is used to reduce the absorption of the atoms to the up-converted light and to avoid pattern formation in the FWM process.

The ladder-type configuration used in our experiment is shown in Fig. 1(a). It consists of one ground state, |1>, two intermediate states, |2> and |3>, and one upper state, |4>, corresponding to the energy levels $5S_{1/2}$, $5P_{3/2}$ and $4D_{5/2}$, respectively. The transition frequency between the energy levels $5S_{1/2}$ and $5P_{3/2}$ corresponds to the $D_2$ line (780 nm) of $^{85}$Rb, and the transition between the $5P_{3/2}$ and $4D_{5/2}$ levels can be coupled using a laser at 1529.4 nm. $\Delta_1$ is the detuning between the $^{85}$Rb atomic transition $5S_{1/2}(F=3)$-$5P_{3/2}(F''=4)$ and the 780 nm laser, and $\Delta_2$ is the detuning of the $^{85}$Rb atomic transition $5P_{3/2}(F''=4)$-$4D_{5/2}(F'''=4)$ with respect to the 1529.4 nm laser. A red frequency shift represents a positive in this report.

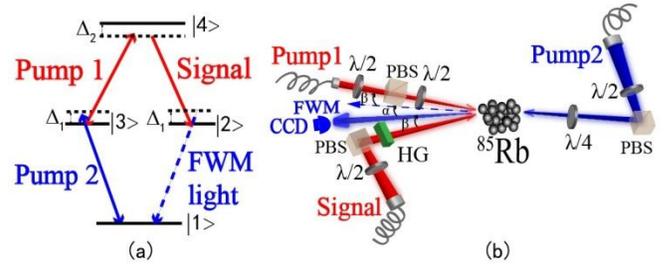

Fig. 1. (Color online) (a) Energy level diagrams of the ladder-type configuration. (b) Schematic diagram of the experimental setup. The red dashed lines indicate the paths of the 1529.4 nm light, and the blue dashed line shows that of the 780 nm light. HG: computer-generated hologram, PBS: polarization beamsplitter, λ/2: half-wave plate, λ/4: quarter-wave plate. α=0.24°, ß=0.37°.

Before introducing our experimental results, we give a simple description of our system. With the assumption that the Rabi frequencies $\Omega_{p1}$, $\Omega_s \gg \Omega_F$, $\Omega_{p2}$, we derive steady-state solutions for the density matrix and obtain the susceptibility of the FWM field using $P_{FWM}=N\mu_{21}\rho_{21}$, where $N$ is the active density of the atoms, and $\mu_{21}$ and $\rho_{21}$ are the dipole element and the density matrix element of the transition |2>->|1> respectively. The evolution of the FWM beam is

described by the following equation:

$$\frac{\partial}{\partial z}\Omega_F - \frac{i}{2k}\nabla^2\Omega_F = \frac{iN\mu_{21}^2 k_F}{2\hbar\varepsilon_0}(A^{(1)}\Omega_F + A^{(2)}\Omega_s^2\Omega_F + A^{(3)}\Omega_{p1}^2\Omega_F + A^{(4)}\Omega_{p1}\Omega_{p2}\Omega_s^*)$$ (1)

where $A^{(1)}$ and $A^{(4)}$ are the loss and gain coefficients corresponding to the absorption and the nonlinear FWM gain, respectively. $A^{(2)}$ and $A^{(3)}$ are the cross-modulation coefficients from the pump 1 and signal fields, respectively. $A^{(i)}$ ($i$=1,2,3,4) are the complex functions of the detuning, Rabi frequency, decay rate, active atomic density and dipole element. If the detuning $\Delta_1$=2 GHz, $\Delta_2$=1.67 GHz is large (our experimental parameters), the sum of the first term, the second term and the third term in brackets on the right of Eq. (1) can be simplified to be $i\Omega_F/2(\gamma_1-i\Delta_1)$, which characterizes the absorption of the FWM light. This illustrates that spatial cross-modulations from pump 1 and the signal light can be avoided by the large detuning in our system. Ref. [16] reports that the spatial cross-modulation can be adjusted through single-photon detuning.

The simplified experimental setup is shown in Fig. 1(b). A cw laser beam at 780 nm from an external-cavity diode laser (DL100, Toptica) is input to an Rb cell as pump 2. A cw laser beam at 1529.4 nm from another external-cavity diode laser (DL100, Prodesign, Toptica) is divided into two beams by a beam splitter to produce the pump 1 and signal beams. These two beams nearly co-propagate and have parallel polarization directions. The angle between these two beams is about 0.98°. The pump 1 beam nearly counter-propagates with the pump 2 beam with an angle of β=0.37°; the pump 2 beam has circular polarization. The up-converted field at 780 nm with the same polarization as the pump 2 beam nearly counter-propagates with the signal with a small angle of α=0.24°, and is monitored using a common CCD camera. The signal and pump 2 beams are focused and the diameters at the center of the Rb cell are approximately 185 $\mu$m and 195 $\mu$m respectively. The pump 1 beam is weakly focused and the beam waist is approximately 1 mm in the cell. A homemade computer-generated hologram (HG) is inserted into the signal beam to generate a light beam carrying the OAM. An HG with a fork structure at the center can be used to change the order of the OAM of a beam. Here, we consider the case where $p$=0, in which the LG$_{0l}$ mode carries the corresponding OAM of $l$. The +/- order diffraction of the HG increases the OAM of the input beam by $\pm\hbar$, when the dislocation of the HG overlaps with the beam center. The superposition of the LG$_{00}$ mode and the LG$_{01}$ mode can be achieved easily by shifting the dislocation of the hologram out of the beam center by a certain amount. The intensity of the LG$_{00}$ mode is Gaussian and the LG$_{01}$ mode has a doughnut-shaped intensity distribution. All experiments were performed using a 5-cm $^{85}$Rb vapor cell, which contains isotopically-pure $^{85}$Rb. After transmission through the HG, the signal beam is divided into two beams: the zero-order diffraction wave and the +-order diffractive wave. The +-order diffractive wave with the doughnut-shaped intensity distribution is incident on the Rb vapor cell as the input signal to be up-converted.

We inserted the HG into the signal beam, and selected the +-order diffractive wave from the HG as the signal beam shown in Fig. 2(a). We used a CCD camera to monitor the generated signal. The powers of the pump 2, pump 1 and signal beams were 0.072 mW, 9.6 mW and 9.6 mW, respectively. The diffractive efficiency of the HG is approximately 10%, so the effective power of the signal was approximately 0.96 mW. The image obtained in the experiment is shown in Fig. 2(b). We also used the superposition of the LG$_{00}$ mode and the LG$_{01}$ mode indicated by Fig. 2(c) as an input, which was obtained by shifting the dislocation of the hologram out of the beam center by a certain amount. The up-converted superpositioned image is shown in Fig. 2(d). Figs. 2(e-h) correspond to the theoretical simulations using Eq. (1).

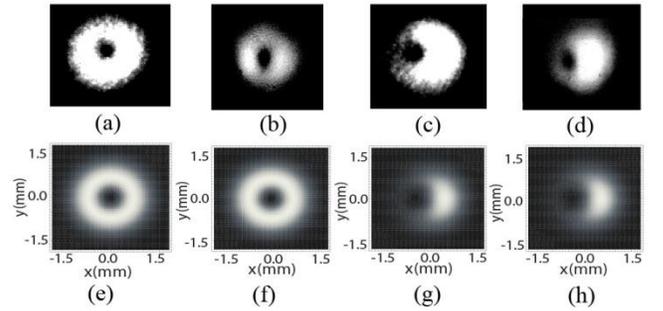

Fig. 2. (a) and (c) show the input multimodes on the signal beams. (b) and (d) show the recorded FWM upconverted images. (e-h) correspond to the theoretical simulation results.

Next, we took a different superposition of OAM modes as the input signal to be up-converted. An arbitrary superposition of the OAM modes αLG$_{00}$+βLG$_{01}$ can be prepared by shifting the dislocation of the hologram out of the beam center by a different amount, where α and β are real without loss of generality, and $|α|^2+|β|^2$=1. The experimental results are shown in Fig. 3a. Figures (a) to (i) on the first line correspond to the superposition of the αLG$_{00}$+βLG$_{01}$ OAM modes with different α and β values, which was realized experimentally by gradually shifting the position of the input beam center relative to the dislocation from one side of the HG to the other. The second line of Fig. 3(a) shows the corresponding up-converted results. We confirmed that the different input signals produced different output signals. We also compared the relative distance ΔX$_{S(F)}$ between the center point of the small black hole and the center point of the entire high intensity shape before and after the experiments. The different relative distance corresponds to the different superposition of the αLG$_{00}$+βLG$_{01}$ OAM modes. The results are shown in Fig. 3(b). We found that the relative distance was maintained linearly. The third and fourth lines correspond to the theoretical simulation results.

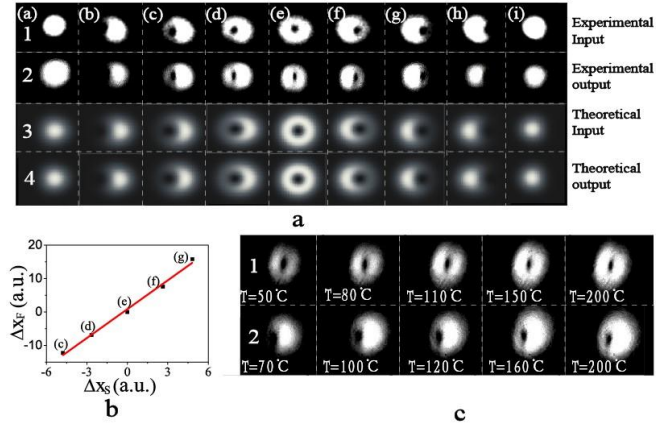

Fig. 3. (Color online) (a). The up-converted image with different input superposition of OAM modes. The first and the second lines show the experimental results, and the third and the fourth lines show the theoretical results. (b). The relative distance between the center point of the small black hole at the center of the intensity shape and the center point of the entire high intensity shape before and after the experiments. The horizontal axis represents the input signal, and the vertical axis represents the up-converted signal. The dots are the experimental data, and the line is the fitted result. Dots (c) to (g) correspond to the experimental cases of (c) to (g) in Fig. 3(a). (c). The up-converted images relative to the cell temperature. In the top line, the input signal is in the $LG_{00}$ mode; in the bottom line, the input signal is the superposition of the $LG_{00}$ mode and the $LG_{01}$ mode.

We also checked the quality of the up-converted image against the cell temperature. We found that, apart from the conversion efficiency, the temperature of the cell has no obvious effect, and it made no difference whether the input signal was a single light beam with OAM or the superposition of two light beams with different OAM modes. The results are shown in Fig. 3(c).

One problem that affects the quality of the transferred image is diffusion caused by atoms in motion during the FWM process. The geometry of our experimental setup causes another problem. The non-collinear configuration is used to reduce the noise from strong input light beams. However, it makes the FWM different along the horizontal and vertical directions.

In conclusion, we have experimentally demonstrated linear up-conversion using the FWM process with large single-photon detuning. Our research results may be useful in research fields including astrophysics, night-vision technology, chemical sensing, and quantum communications.

**Acknowledgments**

DSD and BSS thank Prof. Boyd for the advice about the proof of the linearity, and also thank Dr. X-F Ren and Mr. L-L Wang for providing both the HG and very useful advice. This work was supported by the National Natural Science Foundation of China (Grant Nos. 10874171, 11174271), the National Fundamental Research Program of China (Grant No. 2011CB00200), the Innovation Fund from CAS, and the Program for NCET.